# GaAs valley photonic crystal waveguide with light-emitting InAs quantum dots

Takuto Yamaguchi[1], Yasutomo Ota[2], Ryota Katsumi[1], Katsuyuki Watanabe[2], Satomi Ishida[1], Alto Osada[2], Yasuhiko Arakawa[2], and Satoshi Iwamoto[1,2]

[1]*Institute of Industrial Science, The University of Tokyo, 4-6-1 Komaba, Meguro-ku, Tokyo 153-8505, Japan.*
[2]*Institute for Nano Quantum Information Electronics, The University of Tokyo, 4-6-1 Komaba, Meguro-ku, Tokyo 153-8505, Japan.*

E-mail: takutoym@iis.u-tokyo.ac.jp, iwamoto@iis.u-tokyo.ac.jp

We report a valley photonic crystal (VPhC) waveguide in a GaAs slab with InAs quantum dots (QDs) as an internal light source exploited for the experimental characterization of the waveguide. A topological interface state formed at the interface between two topologically-distinct VPhCs is used as the waveguide mode. We demonstrate robust propagation for the near-infrared light emitted from the QDs even under the presence of sharp bends due to the topological protection of the guided mode. Our work will be of importance for developing robust photonic integrated circuits with small footprints, as well as for exploring active semiconductor topological photonics.

Topology has been a keyword in the field of condensed matter physics[1]. Novel concepts based on topological physics have renewed our understanding of materials and have opened innovative routes to engineer the flow of electrons. Recently, the concept of topology has been extended to photonics, thereby realizing various novel states of electromagnetic waves[2-4]. The hallmark of such topological photonic systems is the existence of topological edge states. They are topologically protected and can guide light with suppressed back reflection even under the presence of defects and sudden changes in the propagation direction. These properties are highly attractive for developing loss-free photonic integrated circuits with ultra-compact footprints.

Topological photonics systems have been implemented on various platforms, including those emulating quantum Hall systems[5] composed of gyromagnetic photonic crystals (PhCs)[6] and waveguide arrays[7], and quantum spin Hall systems consisting of coupled ring resonator arrays[8] and PhCs.[9-14] Among the various topological photonic systems, valley photonic crystals (VPhCs)[15-20] are one of the promising platforms for integrated photonics. They can be realized simply by breaking the spatial inversion symmetry of the lattice as discussed below. Regardless of its simplicity, at the interface between two VPhCs with distinct valley topologies, topologically-protected waveguide modes can be utilized without applying an external magnetic field. The waveguide mode exhibits efficient guiding of light even with sharp turns. Robust light guiding in VPhC waveguide has been experimentally demonstrated at microwave and terahertz frequencies[16,20]. Very recently, aiming at the application to integrated photonics, silicon-based VPhCs slabs that function for transverse electric (TE) polarization at the near infrared region, have been reported[21,22]. Silicon is an excellent material for passive photonic circuits. On the other hand, the use of compound semiconductors is also advantageous because it is possible to introduce efficient active light emitters to VPhC systems. These light emitters can not only be utilized as internal light sources to characterize the valley-protected waveguide modes but are also imperative for realizing active topological photonic devices[23-28].

In this letter, we experimentally demonstrate a VPhC waveguide made of a GaAs slab embedding light-emitting InAs quantum dots (QDs)[29]. The waveguide is formed at the interface between two topologically-distinct slab-type VPhCs, supporting an interface waveguide mode for TE-polarized light. We verified the light transmission through the topological interface mode using photoluminescence from the QDs in the near infrared. The wave propagation was observed to be robust against sharp bends, which is a hallmark of the expected valley protection. Our VPhC waveguide is all-dielectric and planer, and embeds active materials and supports TE-like modes. These properties are desirable to accommodate the demands from photonic integrated circuit technology as well as for exploring active topological photonics.

VPhCs can be constructed from a photonic graphene, a honeycomb PhC with the same refractive index modulation at each sublattice. The presence of both time and spatial inversion symmetry in photonic graphene leads to the formation of photonic Dirac points in the photonic band structure at the equivalent K/K' points[30]. The introduction of different index modulations at two sublattices breaks the spatial inversion symmetry of the system and lifts the degeneracy, leading to the formation of energy gaps and concomitant valleys at K and K'. These valleys are not identical anymore and have opposite Berry curvatures around K and K'[16,21,22]. Another VPhC structure can be created by interchanging the index modulations between two sublattices in the original VPhC. The signs of the Berry curvatures at around K and K' in this VPhC are flipped from those in the original. Thus, these two VPhCs have distinct valley topologies. When interfacing these two VPhCs, the interface supports an edge state as a consequence of bulk-edge correspondence. As discussed in previous works[15-22], the edge state can be a channel for light that is robust against certain types of defects including sharp bends of the interface because the K/K' valleys are well separated from each other in the reciprocal space, and therefore the inter-valley scattering is largely suppressed.

Our VPhC design is based on a honeycomb lattice with a period $a$. Its unit cells are termed as A and B, and are shown in Fig. 1(a). Equilateral triangular air holes with different side lengths are located at different sublattices of the honeycomb lattice. This breaks the spatial inversion symmetry of the original honeycomb lattice. The large and small triangles have side lengths of $1.3 \times a/\sqrt{3}$ and $0.7 \times a/\sqrt{3}$, respectively. In essence, the two unit cells are the same, as the A unit cell coincides with B by flipping upside down. As such, bulk VPhCs based on them share the same photonic band structure as plotted in Fig. 1(b), which is calculated for TE-like modes using the two-dimensional plane wave expansion (2D-PWE) method by assuming the effective refractive index of the slab to be 2.855. Between the first and second bands, a band gap opens with its energy maxima and minima being occupied by the peaks of the K and K' valleys. However, these two structures have topologically different characteristics. The field intensity and phase distributions of the magnetic field perpendicular to the slab plane for the lowest band at K for the structures with unit cells A and B are shown in Fig. 1(c). The distributions of the two structures differ not only in their intensity profiles but also in how the phase distributions rotate in space. More importantly, the latter indicates that the two structures have distinct valley topologies[15-17,22]. Accordingly, the interfaces constituted of the two different VPhCs support waveguide modes originating from the difference in band topology. The band structure projected along the zigzag interface formed by the two VPhCs with $a = 330$ nm and a slab thickness of 200 nm is shown in Fig. 1(d). The band structure is calculated by using the 3D-PWE method. Here, we set the refractive index of the slab as 3.4, which corresponds to the refractive index of GaAs that we used in the following experiments. Two different zigzag interfaces, A-B and B-A interfaces (see Fig. 1(e)) can be formed. In both configurations, topology-induced waveguide modes exist as plotted in Fig. 1(d), where the solid orange (green) curve shows the dispersion for the mode confined at the A-B (B-A) interface. In contrast to the waveguides formed in topological PhCs based on the band folding scheme[9-11], the investigated VPhC waveguide modes exist below the light line, which is beneficial for low-loss light transport and thus for photonic integrated circuit applications. The waveguide modes are localized around the interfaces, as shown in their calculated field profiles plotted in Fig. 1(e). The A-B (B-A) interface supports a symmetric (anti-symmetric) in-plane electric field distribution. In the following, we focus on the A-B interface, as their symmetric field profile is more suitable for photonic integrated circuit applications.

Figure 2 (top panel) shows the transmission spectra for a straight VPhC waveguide and a Z-shaped VPhC waveguide (see the inset for the schematic of the Z-shaped one), calculated by 3D-FDTD. The distance from the waveguide entrance to the detection point in the waveguide is $93a$ for each calculation. The transmittance is defined as the ratio of the detected power to the power emitted from a point-like source at outside the waveguide entrance at a distance of $1a$. Therefore, the insertion loss to the waveguide is included in the result. The single-mode operation band, obtained from the band calculation using 3D-FDTD for consistency, is shaded in yellow. The dashed vertical line indicates the wavelength $\lambda_L$ at which the dispersion curve of the waveguide mode crosses the light line. As long as the operation wavelength is longer than $\lambda_L$, there is no significant degradation of the transmittance in the Z-shaped VPhC waveguide from that in the straight one over the single-mode operation band. This demonstrates the robustness of the waveguide against the sharp bends thanks to the topological protection. Interestingly, for both VPhC waveguides, a high transmittance is observed within the single-mode operation band, even in the leaky region shorter than $\lambda_L$. For comparison, we calculated transmission spectra for a conventional W1 PhC waveguide with straight and Z-shaped propagation channels as plotted in the bottom panel of Fig. 2. The W1 waveguide is patterned in a triangular-lattice GaAs photonic crystal slab (thickness = 200 nm), composed of circular air holes with a diameter of 198 nm. Its lattice constant is tuned to be 300 nm so that its transmission band lies at a similar spectral region as that of the VPhC waveguides. The results show a significant reduction in the transmission when introducing the bends, making a stark contrast with the case of the VPhC waveguide.

We fabricated the designed VPhC waveguides on a GaAs slab with a nominal thickness of 200 nm. InAs QDs which act as internal light emitters, were inserted at the middle of the slab layer. We patterned the structures using conventional semiconductor processing based on electron beam lithography. A scanning electron microscope (SEM) image of a completed sample is shown in Fig. 3(a). The VPhC interface is $70a$ long and is terminated by the semicircle grating ports that out-couple the photons from the waveguide into free space. We optically characterized the samples by micro-photoluminescence (μPL) measurements. The samples were kept in a continuous-flow liquid-helium cryostat and cooled down to around 8 K. We pumped the sample using an 808-nm CW laser diode through an objective lens and utilized PL emission from the QDs as an internal light probe. The PL from the QD ensemble spreads from 920 to 1250 nm due to inhomogeneous broadening and the contributions from the excited states. We analyzed the collected PL signals by using a spectroscope and an InGaAs camera.

First, we investigated the light transmission in a straight interface composed of the A-B domain wall (we name this configuration as "non-trivial"). Such an interface is shown in the SEM image in Fig. 3(a). We excited the QDs embedded in the topological waveguide with a pump power of 5 μW and measured the PL image with a band-pass filter (BPF) of $1050 \pm$ 12.5 nm. The obtained image shown in Fig. 3(b) exhibits bright radiation from the output ports, indicating light propagation in the topological interface. The cross in the image indicates the excitation position. For comparison, we also measured the PL from a device with an A-A interface, which formed a trivial 'waveguide'. In this case, the valley-based propagation mode becomes absent, leading to the disappearance of light emission at the output ports, as shown in Fig. 3(c). We further characterize the topological waveguide mode at the A-B interface by measuring its spectrum at one of the output ports, as plotted in Fig. 3(d). A comparison between the non-trivial and trivial samples shows an increase in the out-coupling of light in the wavelength range from 1.02 to 1.13 µm, which can be considered as the transmission band of the topological mode within the fabricated VPhC interface. However, the propagation band is largely shifted to shorter wavelengths, which arises from thinner slab thickness of ~185 nm (measured by SEM) for the sample used in the experiments than the nominal value. The single-mode operation band and $\lambda_L$, which were calculated using the actual slab thickness, are indicated by the yellow shaded region and the vertical dashed line, respectively, in Fig. 3(d). The spectral range where the non-trivial sample shows stronger emission agrees well with the single-mode operation band. We note that the increased emission is also observed at around 1,180 nm. This could be attributed to the light guiding via the topological waveguide mode, which is overlapping with the bulk band in wavelength (see Fig. 1(d)). We also confirm the spatial localization of the topological mode at the vicinity of the interface by measuring the dependences of the pump position on the PL intensities across the waveguide, as depicted in Fig. 3(e). We monitored the PL intensities at the output port with band-pass filtering for $1,050 \pm 12.5$ nm, while scanning the pump spot along the A-B-C line displayed in Fig. 3(a). We observed a sharp peak at the position of the interface of the device with the topological A-B interface, elucidating the spatial confinement of the mode.

Next, we study light propagation in a Z-shaped VPhC waveguide composed of the A-B interface. An SEM image of a corresponding sample is shown in Fig. 4(a). First, we changed the excitation position and measured the emission spectrum above the output port located at the bottom right. The light-blue curve in Fig. 4(b) shows the emission spectrum that was obtained when the QDs buried around the upper left grating port were excited. Again, we observed an emission peak around 1,070 nm, roughly coinciding with that for the case of the straight waveguide. Note that the spectral region showing strong emission from the output port agrees well again with the calculated single-mode operation band. We also recorded the spectra above the output grating while varying the pump position, as overlaid in Fig. 4(b). A strong emission from the output was observed only when we excited the QD at around the input grating port (position I). These results confirm that the emission peak originates from the QD emission guided through the Z-shaped interface by the topological waveguide mode. Then, we took PL images with BPFs inserted in front of the InGaAs camera, as shown in Fig. 4(c). The left image was taken with a 10-nm-width BPF with a center wavelength of 1,064 nm, which falls into the center of the transmission peak. In this case, we observed strong light radiation from the exit port. More importantly, we did not observe any significant light scattering at the positions of the sharp bends. As expected, such waveguiding effect vanishes when measuring at wavelengths outside the transmission band (see the right panel, taken with a BPF centered at 1,150 nm with a 25-nm bandwidth). These results experimentally demonstrate the robust waveguiding in the VPhC waveguide.

In conclusion, we demonstrated light propagation through a GaAs VPhC waveguide operating in the near infrared by utilizing the emission from InAs QDs embedded in the waveguide. The waveguide is formed at the interface between two topologically-distinct VPhCs. We numerically and experimentally demonstrate robust waveguiding in the structure even under the presence of sharp bends. Our waveguide is based on a thin GaAs slab embedding InAs QDs and supports the propagation of TE-like modes, which are advantageous for photonic integrated circuit applications especially with active semiconductor materials. We envision that the VPhC platform presented here will be a stepping stone for exploring both passive and active topological photonics.

**Acknowledgments**

The authors would like to acknowledge Ingi Kim, the University of Tokyo and Yasuhiro Hatsugai, University of Tsukuba for fruitful discussions. This work was partially supported by MEXT KAKENHI Grant Number JP15H05700, JP15H05868, and JP17H06138.

## Figure Captions

**Fig. 1.** (a) Schematics of the unit cells of the VPhCs. (b) Bulk photonic band diagram for TE-like modes. The gray regions indicate the light cone. The inset shows the first Brillouin zone. (c) Intensity and phase distributions of the modes of the 1st bulk bands at K. (d) Projected photonic band diagrams for the A-B and B-A interfaces. The gray (light gray) shaded regions represents the bulk bands (light cone). (e) Electric field distributions for the topological waveguide modes computed at $k_x a / 2\pi = 1/3$.

**Fig. 2.** Simulated transmission spectra for the VPhC and W1-PhC waveguides. The yellow regions indicate the single-mode operation band calculated using 3D-FDTD. The dashed vertical lines indicate the longest wavelength edge of the leaky region.

**Fig. 3.** (a) SEM image of a straight VPhC waveguide with the A-B zig-zag interface. The insets show magnified images around the domain boundary, together with that for a trivial interface. Measured PL images of (b) non-trivial and (c) trivial samples. (d) μPL spectra measured via an output grating when the QDs are excited at point B indicated in (a). (e) PL intensities measured via the output grating at various excitation positions (from A to C). The intensities are averaged values over 1063−1068 nm.

**Fig. 4.** (a) SEM image of a Z-shaped VPhC waveguide. The inset shows a magnified view of one of the 120-degree bends. (b) PL spectra measured via the grating port by exciting the various locations defined in (a). (c) PL images measured by exciting the position I with two different BPFs as described in the main text. The red dotted lines and yellow circles respectively indicate the periphery of the VPhC and the positions of the output gratings.

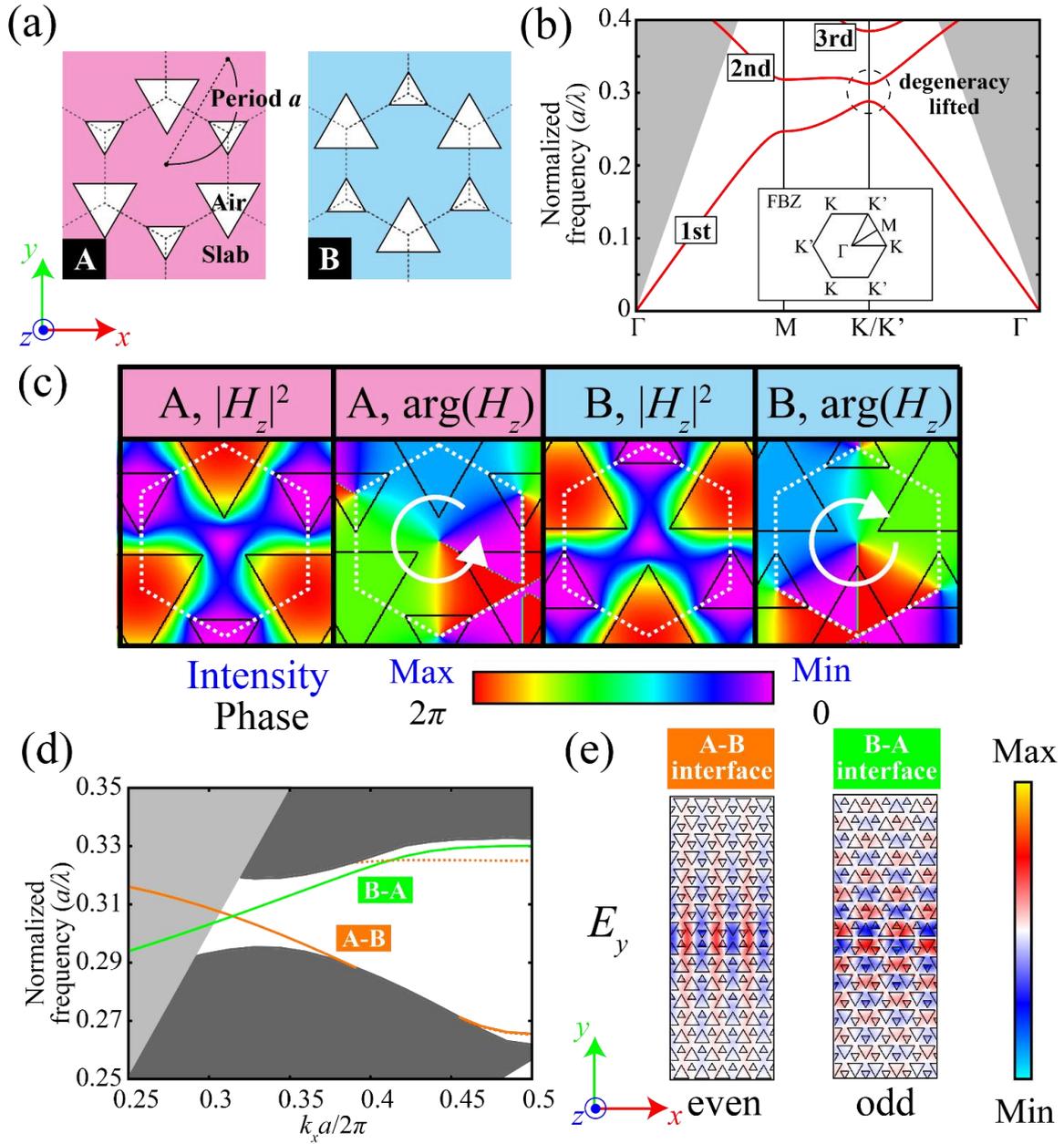

Fig.1.

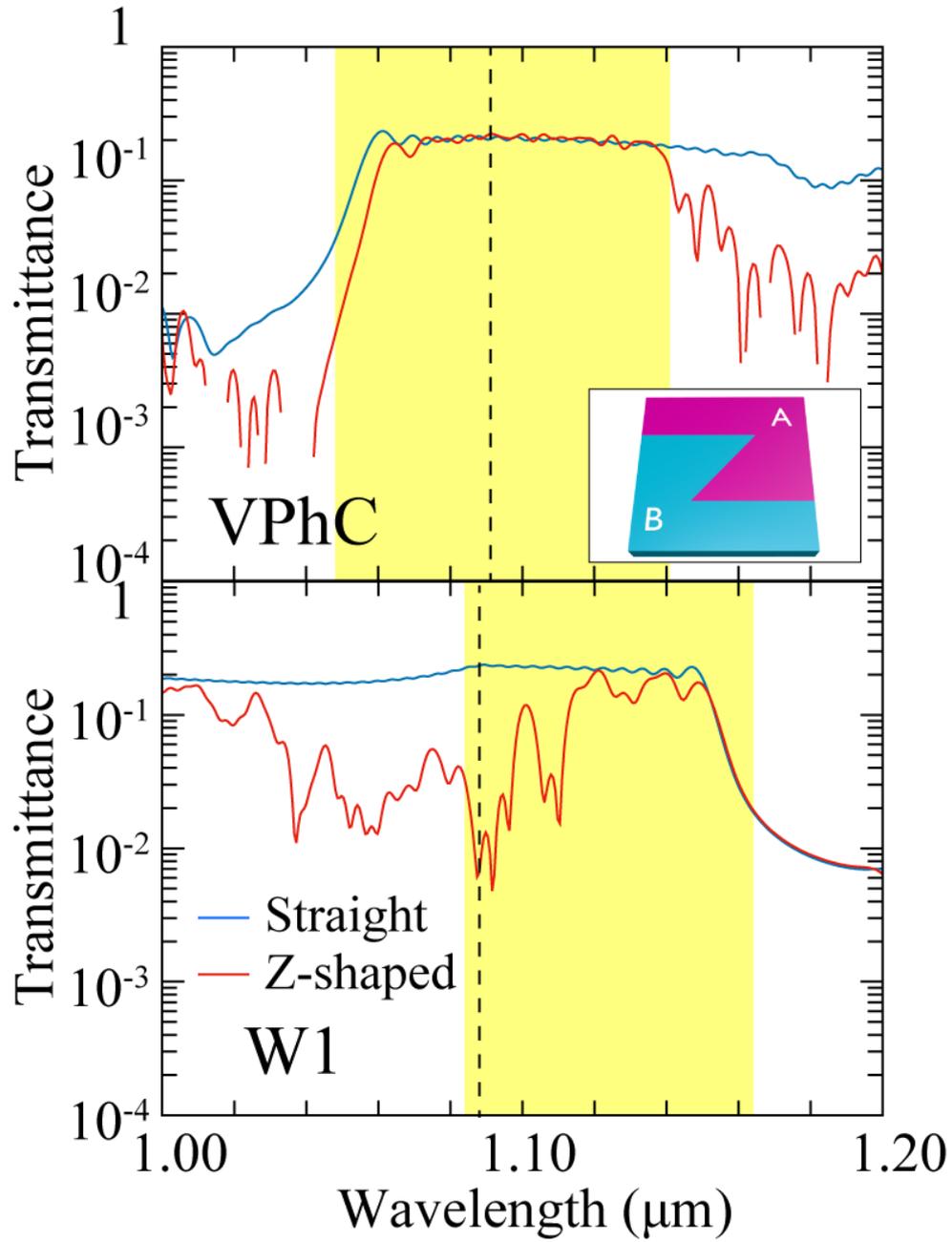

Fig.2.

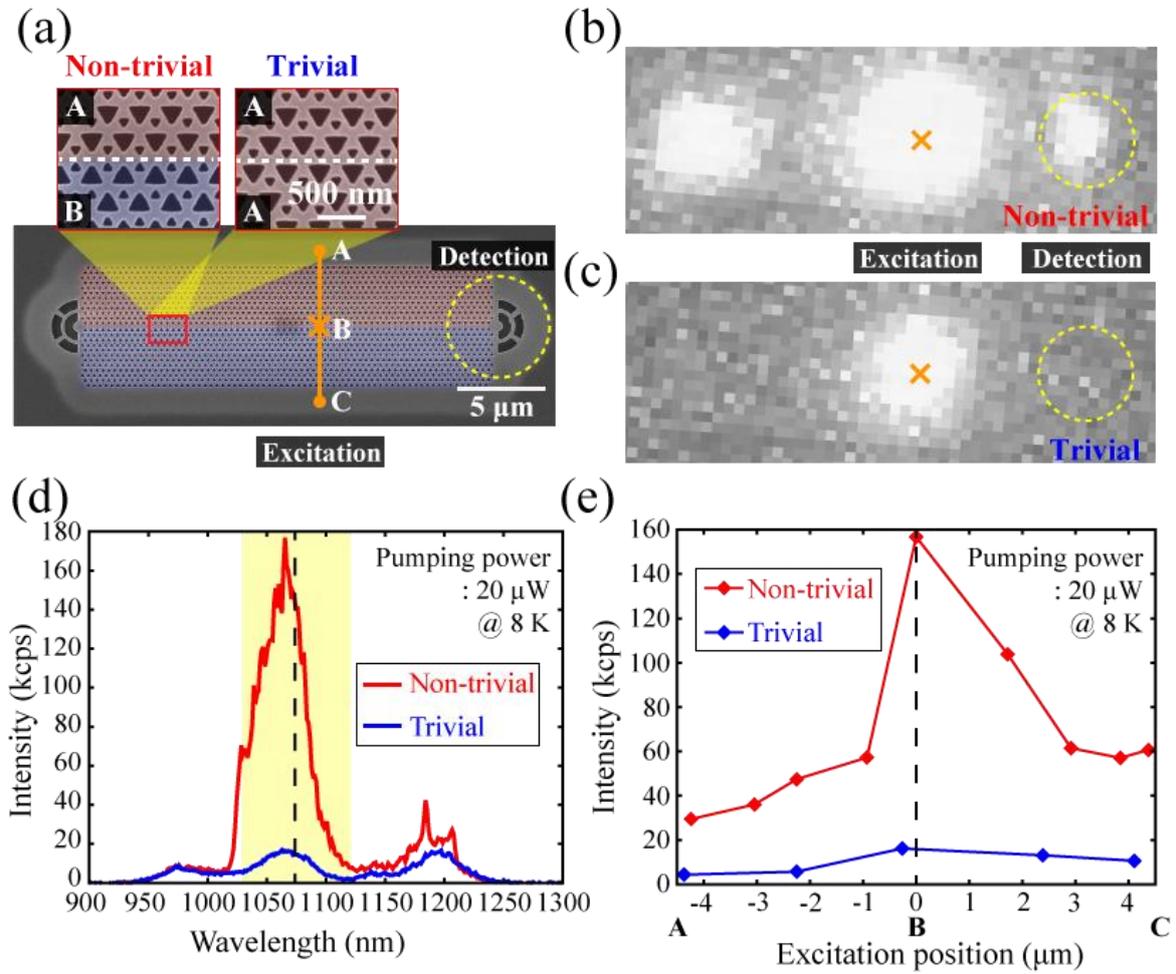

Fig.3.

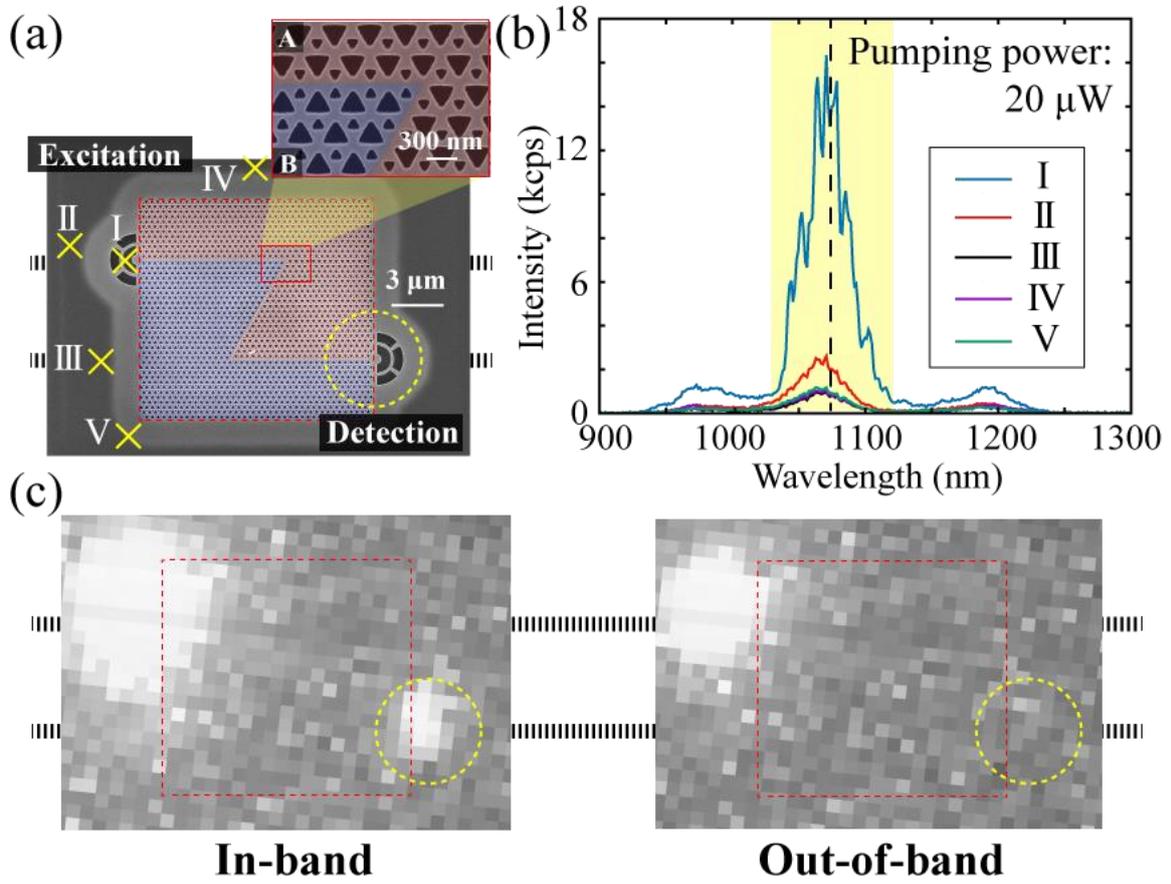

Fig.4.